\documentclass[aps,prd,superscriptaddress,showkeys,showpacs]{revtex4}
\usepackage[T1]{fontenc}
\usepackage{mathptmx}
\usepackage{epsfig,epsf}
\usepackage{amsmath}
\usepackage{amsthm}
\usepackage{amsfonts}
\usepackage{amssymb}
\usepackage{dsfont}

\usepackage{slashed}

\usepackage[active]{srcltx}

\newcommand{\be}{\begin{equation}}
\newcommand{\ee}{\end{equation}}
\newcommand{\ba}{\begin{eqnarray}}
\newcommand{\ea}{\end{eqnarray}}
\newcommand{\baa}{\begin{eqnarray*}}
\newcommand{\btab}{\begin{tabular}}
\newcommand{\etab}{\end{tabular}}
\newcommand{\eaa}{\end{eqnarray*}}

\def\inbar{\,\vrule height1.5ex width.4pt depth0pt}
\def\IC{\relax\hbox{$\inbar\kern-.3em{\rm C}$}}
\def\IZ{\relax{\hbox{\cmss Z\kern-.4em Z}}}
\def\IR{{\hbox{{\rm I}\kern-.2em\hbox{\rm R}}}}

\def\IP{{\hbox{{\rm I}\kern-.2em\hbox{\rm P}}}}
\def\II{\hbox{{1}\kern-.25em\hbox{l}}}

\begin{document}

\title{NLO light-cone sum rules for the nucleon electromagnetic form factors}

\author{I.V.~Anikin}
\affiliation{Bogoliubov Laboratory of Theoretical Physics, JINR, 141980 Dubna, Russia}
\author{V.M.~Braun}
\affiliation{Institut f\"ur Theoretische Physik, Universit\"at
   Regensburg,D-93040 Regensburg, Germany}
\author{N.~Offen}
\affiliation{Institut f\"ur Theoretische Physik, Universit\"at
   Regensburg,D-93040 Regensburg, Germany}
\date{\today}
\begin{abstract}
  \vspace*{0.3cm}
\noindent
We study the electromagnetic nucleon form factors within the approach based on
light-cone sum rules. We include
the next-to-leading-order corrections for the contributions of twist-three and twist-four
operators and a consistent treatment of the nucleon mass corrections in our calculation.
It turns out that a self-consistent picture arises when the three valence quarks
carry $40\%:30\%:30\%$ of the proton momentum.
 \end{abstract}
\maketitle
\date{\today}
\section{Introduction}

We derive light-cone sum rules (LCSRs) for the electromagnetic nucleon form factors including
next-to-leading-order corrections for the contribution of twist-three and twist-four
operators and a consistent treatment of nucleon mass corrections.
The soft Feynman contributions are calculated in terms
of small transverse distance quantities using dispersion relations and duality.
The form factors are expressed in terms of nucleon wave functions at small
transverse separations (DAs), without any additional parameters.
The distribution amplitudes can be extracted from
experimental data on form factors and compared to the results of lattice QCD simulations.
A self-consistent picture emerges, with the three valence quarks carrying $40\%:30\%:30\%$
of the proton momentum.

Our work can be split into three essential parts:
(i) calculations within LCSR;
(ii) derivation of the factorized amplitude at the leading order (LO) up to twist-6 and
at the next-to-leading order (NLO) up to twist-4. We calculated 22 coefficient functions at NLO and 20 of them are new ones.
To avoid mixing with
the so-called evanescent operators, we use the renormalization procedure for operators with
open Dirac indices;
(iii) study of the corresponding distribution amplitudes. In particular, the
light-cone expansion to twist-4 accuracy of the three-quark matrix elements with
generic quark positions.

\section{LCSRs for nucleon form factors: General structure}

The LCSR approach allows one to calculate the form factors in terms of the nucleon (proton) DAs.
To this end we consider the correlation function
\begin{equation}
\label{correlator}
T_{\nu}(P,q) = i\! \int\! d^4 x \, e^{i q x}
\langle 0| T\left[\eta(0) j_{\nu}^{\mathrm{em}}(x)\right] |P\rangle
\end{equation}
where $\eta(0)$ is the Ioffe interpolating current:$
\eta(x) = \epsilon^{ijk} \left[u^i(x) C\gamma_\mu u^j(x)\right]\,\gamma_5 \gamma^\mu d^k(x)$ and
$\langle 0| \eta(0)|P\rangle  = \lambda_1 m_N N(P)$.
The matrix element of the electromagnetic current
$j_{\mu}^{\rm em}(x) = e_u \bar{u}(x) \gamma_{\mu} u(x) + e_d \bar{d}(x) \gamma_{\mu} d(x)$
taken between nucleon states is conventionally written in terms of the
{Dirac} and {Pauli form factors} $F_1(Q^2)$ and $F_2(Q^2)$:
\begin{eqnarray}
\label{F1F2}
\langle P'| j_{\mu}^{\rm em}(0)|P\rangle =
\bar{N}(P')\left[\gamma_{\mu}F_1(Q^2)-i\frac{\sigma_{\mu\nu}q^{\nu}}{2m_N}F_2(Q^2)\right]N(P).
\end{eqnarray}
In terms of the {electric} $G_E(Q^2)$  and {magnetic} $G_M(Q^2)$
Sachs form factors, we have
\begin{eqnarray}
\label{GMGE}
G_M(Q^2)=F_1(Q^2)+F_2(Q^2),
\quad
G_E(Q^2)=F_1(Q^2)-\frac{Q^2}{4m_N^2}F_2(Q^2).
\end{eqnarray}
We also define a light-like vector $n_\mu$ by the condition
$q\cdot n =0\,,n^2 =0$
and introduce the second light-like vector as
$p_\mu=P_\mu - n_\mu m_N^2/(2 P\cdot n)\,, p^2=0\,,$
and
$g^\perp_{\mu\nu} = g_{\mu\nu}-(p_\mu n_\nu+ p_\nu n_\mu)/(pn)\,$.
We consider the ``plus'' spinor projection of the correlation function
involving the ``plus'' component of the electromagnetic current,
which can be parametrized in terms of two invariant functions
\begin{equation}
\label{project4}
  \Lambda_+ T_+ = p_+\left\{ m_N \mathcal{A}(Q^2,P'^2)  +
    \hat q_\perp \mathcal{B}(Q^2,P'^2)\right\}N^+(P)\,,
\end{equation}
where $Q^2=-q^2$ and $P'^2 = (P-q)^2$ and $N^\pm(P) = \Lambda^\pm N(P),$ $\Lambda^+ = \hat p \hat n/(2 pn)$,
$\Lambda^- = \hat n \hat p/(2 pn)$.
Further, making use of the Borel transformation
$(s-P'^2)^{-1} \longrightarrow e^{-s/M^2}$, one obtains the following sum rules:
\begin{eqnarray}
2\lambda_1 F_1(Q^2)= \frac1{\pi}\int_0^{s_0}ds \, e^{(m_N^2-s)/M^2}
 \text{Im}\, \mathcal{A}^{\rm QCD}(Q^2,s)\,,\quad
\lambda_1 F_2(Q^2) = \frac1{\pi}\int_0^{s_0}ds \,
  e^{(m_N^2-s)/M^2} \text{Im}\, \mathcal{B}^{\rm QCD}(Q^2,s)\,.
\label{eq:LCSRscheme}
\end{eqnarray}
The correlation functions  $\mathcal{A}(Q^2,P'^2)$ and  $\mathcal{B}(Q^2,P'^2)$ can be
written as a sum:
\begin{eqnarray}
\mathcal{A}= e_d\,\mathcal{A}_d + e_u \mathcal{A}_u\,,\quad
\mathcal{B}= e_d\,\mathcal{B}_d + e_u \mathcal{B}_u\,.
\end{eqnarray}
Each of the functions has a perturbative expansion which we write as
\begin{eqnarray}
 \mathcal{A} = \mathcal{A}^{\rm LO}  + \frac{\alpha_s(\mu)}{3\pi}\mathcal{A}^{\rm NLO} +\ldots
\label{AB-NLO}
\end{eqnarray}
and similar for $\mathcal{B}$; $\mu$ is the renormalization scale.
For consistency with our NLO calculation, we rewrite our results in a different
form, expanding all kinematic factors in powers of $m_N^2/Q^2$. We keep all corrections
$\mathcal{O}(m_N^2/Q^2)$ but neglect terms $\mathcal{O}(m_N^4/Q^4)$ etc. which
is consistent with taking into account contributions of twist-three, -four, -five
(and, partially, twist-six) in the operator product expansion (OPE).
%
%
%
After calculations, the NLO corrections read (see all details in \cite{ABO}).
\begin{eqnarray}
&&Q^2{\cal A}^{\rm {NLO}}_q =
\int[dx_i]\biggl\{ \sum_{k=1,3}\Big[\mathbb{V}_k(x_i) C^{\mathbb{V}_k}_q(x_i, W)
+ \mathbb{A}_k(x_i) C^{\mathbb{A}_k}_q(x_i, W)\Big]
\nonumber\\&&{}
+ \sum_{m=1,2,3}\Big[ \mathbb{V}^{(m)}_2(x_i) C^{\mathbb{V}^{(m)}_2}_q(x_i, W)
+\mathbb{A}^{(m)}_2(x_i) C^{\mathbb{A}^{(m)}_2}_q(x_i, W)\Big]
\biggr\}
+ \mathcal{O}(\text{twist-5})
\label{NLO-A}
\end{eqnarray}
and
\begin{eqnarray}
Q^2{\cal B}^{\rm {NLO}}_q=
\int[dx_i] \biggl[
\mathbb{V}_1(x_i) D^{\mathbb{V}_1}_q(x_i, W) +
\mathbb{A}_1(x_i) D^{\mathbb{A}_1}_q(x_i, W)\biggr]
+ \mathcal{O}(\text{twist-5}).
\label{NLO-B}
\end{eqnarray}
Notice that
$C^{\mathbb{V}^{(1)}_2}_d(x_i, W) = C^{\mathbb{A}^{(1)}_2}_d(x_i, W) =0$.
As an example, we present here only the two simplest coefficient functions:
\begin{eqnarray}
&&x_2 C^{\mathbb{V}_1}_d(x_i)=
2x_2 x_3 \Big[ 3 (L-2) g_1(x_3) + 2 (L-1) g_{11}(x_3,x_3) + g_{21}(x_3,x_3)\Big]
 + \Big[2x_2 + (4L-3)x_3\Big] h_{11}(x_3) + (3-4L) \bar x_1  h_{11}(\bar x_1)
\nonumber\\&&{}
+ 2 x_3 h_{21}(x_3) -2 \bar x_1 h_{21}(\bar x_1)
    - 2\Big[3 (x_2/x_3) (2L\!-\!3) + 5L-7 \Big] h_{12}(x_3)
 + 2 (5L\!-\!7) h_{12}(\bar x_1)
    - \Big[6 (x_2/x_3)+ 5 \Big] h_{22}(x_3)
\nonumber\\[1mm]&&{}
+ 5 h_{22}(\bar x_1) + (6/x_3)(L-2) h_{13}(x_3)  - (6/\bar x_1)(L-2) h_{13}(\bar x_1)
+ (3/x_3) h_{23}(x_3) - (3/\bar x_1) h_{23}(\bar x_1)\,,
\nonumber
\end{eqnarray}
and
\begin{eqnarray}
&&x_2 C^{\mathbb{A}_1}_d(x_i)=
3 \bar x_1 h_{11}(\bar x_1)-3 x_3 h_{11}(x_3)
+ 2(3 L-10) h_{12}(\bar x_1)
-2(3L-10) h_{12}(x_3)+3 h_{22}(\bar x_1)-3 h_{22}(x_3) \hspace*{3.5cm}\phantom{.}
\nonumber\\[2mm]
&&
-(6/\bar x_1) (L-3) h_{13}(\bar x_1)
+(6/x_3) (L-3) h_{13}(x_3)
-(3/\bar x_1) h_{23}(\bar x_1)
+(3/x_3) h_{23}(x_3)\,,
\nonumber
\end{eqnarray}
where
\begin{align}
  g_{nk}(y,x;W) = \frac{\ln^n[1-yW-i\eta]}{(-1+ x W +i\eta)^k},
\quad
  h_{nk}(x;W) = \frac{\ln^n[1-x W-i\eta]}{(W +i\eta)^k}
\nonumber
\end{align}
with $n=0,1,2$ and $k=1,2,3$. For $n=0$ the first argument
becomes dummy, i.e
$g_k(x;W) \equiv g_{0k}(\ast,x;W)$.

\section{Results}

\begin{figure*}[t]
   \begin{center}
\includegraphics[width= 5.5cm, clip = true]{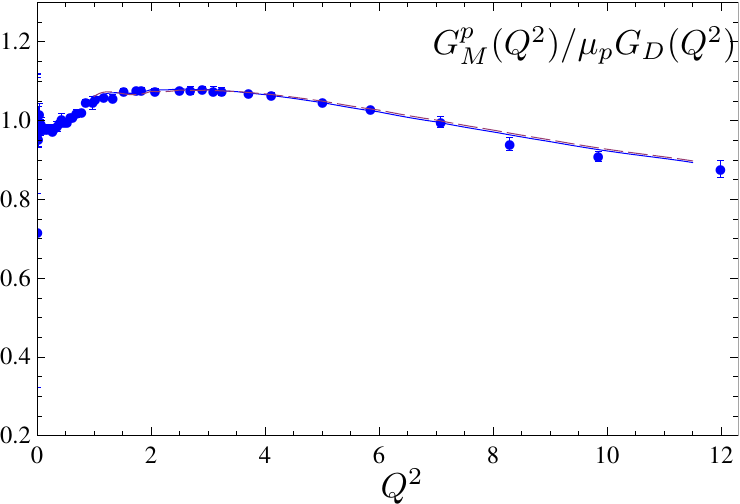}
\qquad
\includegraphics[width= 5.5cm, clip = true]{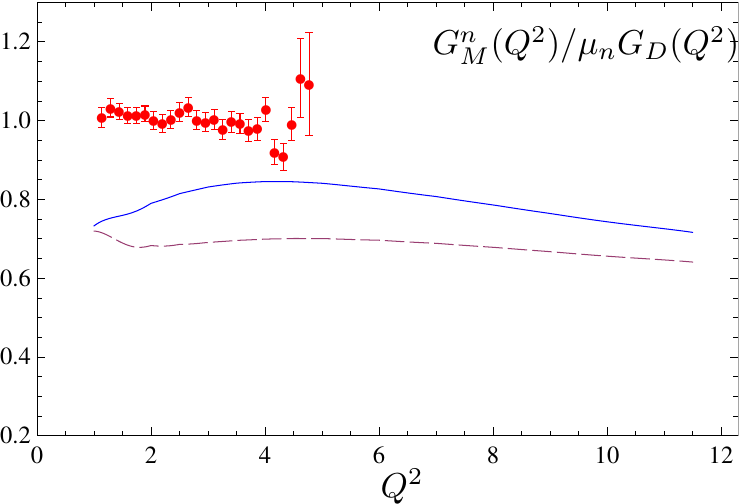}
\\[5mm]
\includegraphics[width= 5.5cm, clip = true]{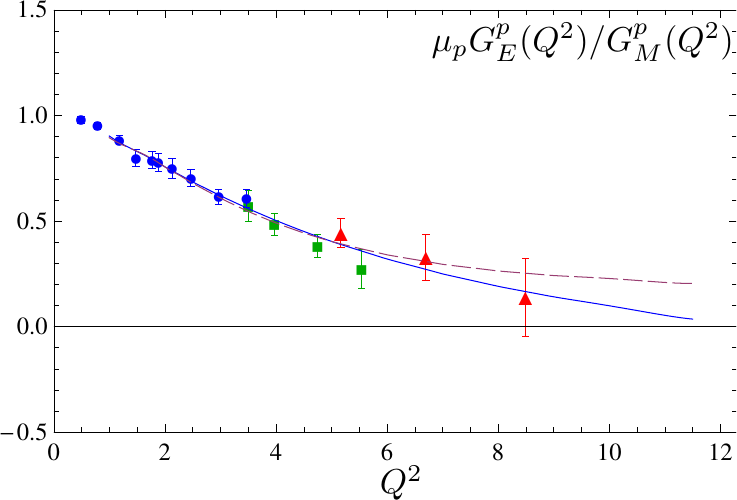}
\hspace*{0.5cm}
\includegraphics[width= 5.5cm, clip = true]{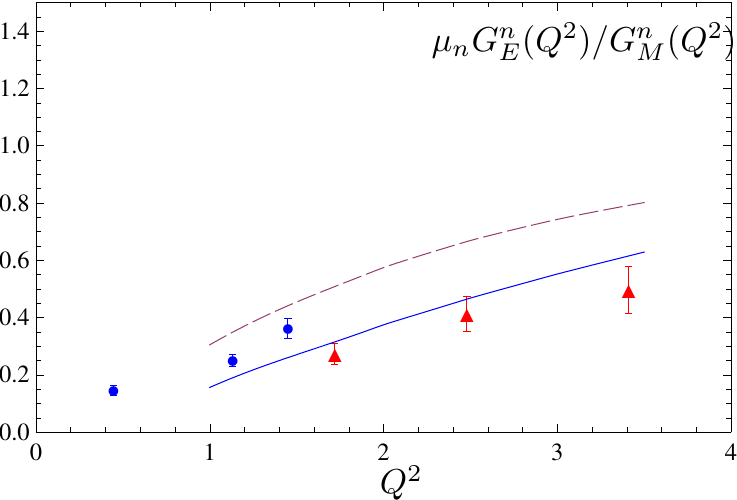}
   \end{center}
\caption{Nucleon electromagnetic form factors from LCSRs compared to the experimental data
\cite{Arrington:2007ux,Lachniet:2008qf,Gayou:2001qd,Punjabi:2005wq,Puckett:2010ac,Plaster:2005cx,Riordan:2010id}.
Parameters of the nucleon DAs correspond to the sets ABO1 and ABO2 in Table~\ref{tab:shape} for the solid and dashed
curves, respectively. Borel parameter $M^2=1.5$~GeV$^2$ for ABO1 and $M^2=2$~GeV$^2$ for ABO2.
}
\label{fig:ABO1}
\end{figure*}

\begin{figure}[ht]
   \begin{center}
\includegraphics[width= 6.5cm, clip = true]{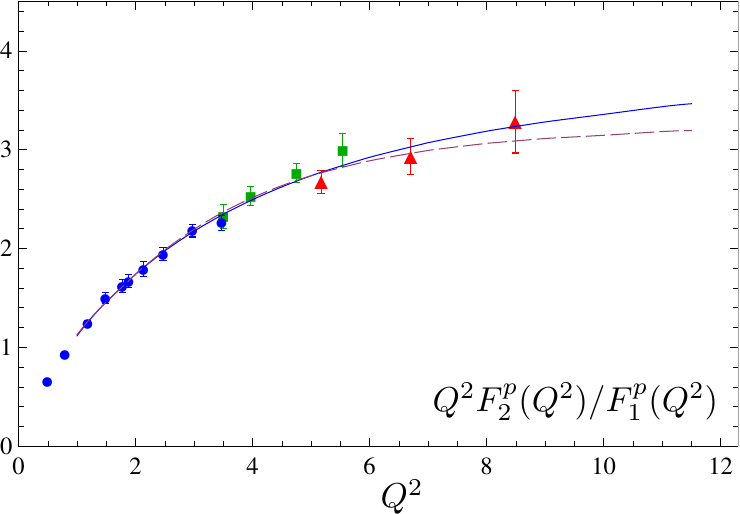}
   \end{center}
\caption{The ratio of Pauli and Dirac electromagnetic proton form factors from LCSRs compared to the experimental data
\cite{Gayou:2001qd,Punjabi:2005wq,Puckett:2010ac}.
Parameters of the nucleon DAs correspond to the sets ABO1 and ABO2 in Table~\ref{tab:shape} for the solid and dashed
curves, respectively. Borel parameter $M^2=1.5$~GeV$^2$ for ABO1 and $M^2=2$~GeV$^2$ for ABO2.}
\label{fig:ABO1-F2F1}
\end{figure}

In this section, we discuss very shortly our main results. The full and
comprehensive analysis and discussion of all input parameters, form factors and DAs
can be found in \cite{ABO}. It is instructive to write down schematically the
structure of all our form factors as
\begin{eqnarray}
{\cal F}={\cal F}_0+\frac{f_N}{\lambda_1}{\cal F}_{f_N}+
\sum\limits_{i=0,1}\eta_{1i}{\cal F}_{\eta_{1i}} +
\frac{f_N}{\lambda_1} \sum\limits_{i=1}^{2}\sum\limits_{j=0;j\leq i}^{2}
\varphi_{ij}{\cal F}_{\varphi_{ij}}\,.
\end{eqnarray}
Main nonperturbative input in the LCSR calculation of form factors is provided by the
normalization constants, $f_N,\, \lambda_1$, and shape parameters of nucleon DAs, $\varphi_{ij}$ and $\eta_{ij}$.
The existing information, together with our final choices explained below, is summarized in Table~\ref{tab:shape}.
As it is seen from Table~\ref{tab:shape}, there only exist quantitative estimates
for $f_N/\lambda_1$ and the first-order shape parameters
$\varphi_{10}$, $\varphi_{11}$ of the leading twist-3 DA. The other parameters, in contrast,
are very weakly constrained. Experimental data favors larger values of $f_N/\lambda_1$
 so that we fix $f_N/\lambda_1=-0.17$
and also take $\varphi_{10}=\varphi_{11} = 0.05$ in agreement with lattice calculations and
the previous LO LCSR studies~\cite{Braun:2006hz}. We then make a fit to the experimental
data on the magnetic proton form factor $G_M^p(Q^2)$ and
$G_E^p/G_M^p$ in the interval $1 < Q^2 < 8.5$~GeV$^2$ with all other entries as free parameters.
We did two separate fits for $M^2=1.5$~GeV$^2$ and  $M^2=2$~GeV$^2$
that are referred as ABO1 and ABO2, respectively. The resulting values for the shape parameters
are collected in Table~\ref{tab:shape} and the corresponding form factors (solid curves for the set ABO1 and dashed for ABO2) are
shown in Fig.~\ref{fig:ABO1} for the proton (left two panels) and the neutron (right two panels). 
The ratio $Q^2 F^p_2(Q^2)/F^p_1(Q^2)$ of Pauli and Dirac form factors in the proton is demonstrated in Fig.~\ref{fig:ABO1-F2F1}. 
The quality of the two fits of the proton data is roughly similar, whereas the description of the neutron form factors is slightly worse for ABO2 compared to ABO1. 
In both fits the neutron magnetic form factor comes out to be 20-30\% below the data.

\begin{table*}[t]
\begin{center}
\begin{tabular}{@{}l|l|l|l|l|l|l|l|l|l|l@{}} \hline
Model &  Method    & $f_N/\lambda_1 $ & $\varphi_{10}$ & $\varphi_{11}$ & $\varphi_{20}$ & $\varphi_{21}$ & $\varphi_{22}$ & $\eta_{10}$   & $\eta_{11}$    & Reference \\ \hline
ABO1  & LCSR (NLO) & $-0.17$         & $0.05$        & $0.05$        & $0.075(15)$   & $-0.027(38)$ & $0.17(15)$    & $-0.039(5)$ & $0.140(16)$   & this work \\ \hline
ABO2  & LCSR (NLO) & $-0.17$         & $0.05$        & $0.05$        & $0.038(15)$   & $-0.018(37)$ & $-0.13(13)$   & $-0.027(5)$ & $0.092(15)$   & this work \\ \hline
BLW   & LCSR (LO)  & $-0.17$         & $0.0534$      & $0.0664$      & -             & -            & -             & $0.05$      & $0.0325$      & \cite{Braun:2006hz}    \\ \hline
BK    & pQCD       & -               & $0.0357$      & $0.0357$      & -             & -            & -             & -           & -             & \cite{Bolz:1996sw}     \\ \hline
COZ   & QCDSR (LO) & -               & $0.163$       & $0.194$       & $0.41$        & $\phantom{-}0.06$ & $-0.163$ & -           & -             & \cite{Chernyak:1987nv} \\ \hline
KS    & QCDSR (LO) & -               & $0.144$       & $0.169$       & $0.56$        & $-0.01$      & $-0.163$      & -           & -             & \cite{King:1986wi}     \\ \hline
HET   & QCDSR (LO) & -               & $0.152$       & $0.205$       & $0.65$        & $-0.27$      & $\phantom{-} 0.020$ & -     & -             & \cite{Stefanis:1992nw} \\ \hline
      & QCDSR (NLO)& $-0.15$         & -             & -             & -             & -            & -             & -           & -             & \cite{Gruber:2010bj}   \\ \hline
LAT09 & LATTICE    & $-0.083(6)$     & $0.043(15)$   & $0.041(14)$   & $\phantom{-} 0.038(100)$ & $-0.14(15)$  & $-0.47(33)$ & -  & -             & \cite{Braun:2008ur}    \\ \hline
LAT13 & LATTICE    & $-0.075(5)$     & $0.038(3)$    & $0.039(6)$    & $-0.050(80)$  & $-0.19(12)$  & $-0.19(14)$   & -           & -             & \cite{lattice2013}     \\ \hline
\end{tabular}
\end{center}
\caption[]{\sf Parameters of the nucleon distribution amplitudes at the scale $\mu^2=2$~GeV$^2$.
 For the lattice results \cite{lattice2013} only statistical errors are shown.
 }
\label{tab:shape}
\renewcommand{\arraystretch}{1.0}
\end{table*}

%


\section{Conclusions}

In conclusion, our calculation incorporates the following
new elements as compared to previous studies:
(i) NLO QCD corrections to the contributions of twist-three and twist-four DAs;
(ii) the exact account of ``kinematic'' contributions to the nucleon DAs of twist-four and twist-five
induced by lower geometric twist operators (Wandzura-Wilczek terms);
(iii) the light-cone expansion to twist-four accuracy of the three-quark matrix
elements with generic quark positions;
(iv) a new calculation of twist-five off-light cone contributions;
(v) a more general model for the leading-twist DA, including contributions of second-order polynomials.

\vspace{0.5cm}
This work was supported by the German Research Foundation (DFG), grant BR 2021/6-1 and
in part by the RFBR (grant 12-02-00613) and the Heisenberg-Landau Program.


\begin{thebibliography}{99}
\bibitem{ABO}
  I.~V.~Anikin, V.~M.~Braun and N.~Offen,
  Phys.\ Rev.\ D {\bf 88}, 114021 (2013)

\bibitem{Braun:2006hz}
  V.~M.~Braun, A.~Lenz and M.~Wittmann,
  Phys.\ Rev.\ D {\bf 73}, 094019 (2006)

\bibitem{Bolz:1996sw}
  J.~Bolz and P.~Kroll,
  Z.\ Phys.\ A {\bf 356}, 327 (1996).

\bibitem{Chernyak:1987nv}
  V.~L.~Chernyak, A.~A.~Ogloblin and I.~R.~Zhitnitsky,
  Z.\ Phys.\ C {\bf 42}, 583 (1989).
\bibitem{King:1986wi}
  I.~D.~King and C.~T.~Sachrajda,
  Nucl.\ Phys.\ B {\bf 279}, 785 (1987).

\bibitem{Stefanis:1992nw}
  N.~G.~Stefanis and M.~Bergmann,
  Phys.\ Rev.\ D {\bf 47}, 3685 (1993);
  N.~G.~Stefanis,
  Eur.\ Phys.\ J.\ direct C {\bf 7}, 1 (1999)

\bibitem{Gruber:2010bj}
  M.~Gruber,
  Phys.\ Lett.\ B {\bf 699}, 169 (2011).

\bibitem{Braun:2008ur}
  V.~M.~Braun {\it et al.}  [QCDSF Collaboration],
  Phys.\ Rev.\ D {\bf 79}, 034504 (2009).
\bibitem{lattice2013}
 R.~Schiel {\it et al.}, \emph{Wave functions of the Nucleon and the $N^\ast(1535)$},
 invited talk at the 31st International Symposium on Lattice Gauge Theory,
 July 29 -- August 03 (2013), Mainz, Germany.

\bibitem{Arrington:2007ux}
  J.~Arrington, W.~Melnitchouk and J.~A.~Tjon,
  Phys.\ Rev.\ C {\bf 76}, 035205 (2007).
\bibitem{Lachniet:2008qf}
  J.~Lachniet {\it et al.}  [CLAS Collaboration],
  Phys.\ Rev.\ Lett.\  {\bf 102}, 192001 (2009).
\bibitem{Gayou:2001qd}
  O.~Gayou {\it et al.}  [Jefferson Lab Hall A Collaboration],
  Phys.\ Rev.\ Lett.\  {\bf 88}, 092301 (2002).
\bibitem{Punjabi:2005wq}
  V.~Punjabi {\it et al.},
  Phys.\ Rev.\ C {\bf 71}, 055202 (2005)
  [Erratum-ibid.\ C {\bf 71}, 069902 (2005)].
\bibitem{Puckett:2010ac}
  A.~J.~R.~Puckett {\it et al.},
  Phys.\ Rev.\ Lett.\  {\bf 104}, 242301 (2010).

\bibitem{Plaster:2005cx}
  B.~Plaster {\it et al.}  [Jefferson Laboratory E93-038 Collaboration],
  Phys.\ Rev.\ C {\bf 73}, 025205 (2006).
\bibitem{Riordan:2010id}
  S.~Riordan {\it et al.},
  Phys.\ Rev.\ Lett.\  {\bf 105}, 262302 (2010).

\end{thebibliography}
\end{document}